\documentclass[aps,preprint,showpacs,superscriptaddress,groupedaddress]{revtex4}  

\usepackage[latin1]{inputenc}
\usepackage{epsfig}
\usepackage{enumerate}
\usepackage{subfigure}
\usepackage{epstopdf}
\usepackage{color}

\begin{document}

\title{Conway's game of life is a near-critical metastable state in the multiverse of cellular automata}
\author{Sandro M. Reia}
\affiliation{Faculdade de Filosofia, Ciências e Letras de Ribeirão Preto, Universidade de São Paulo, Ribeirão Preto, São Paulo, Brazil}
\author{Osame Kinouchi}
\affiliation{Faculdade de Filosofia, Ciências e Letras de Ribeirão Preto, Universidade de São Paulo, Ribeirão Preto, São Paulo, Brazil}

\begin{abstract}

	Conway's cellular automaton Game of LIFE has been conjectured to be a critical (or quasicritical) dynamical system. 
	This criticality is generally seen as a continuous order-disorder transition in cellular automata (CA) rule space. 
	LIFE's mean-field return map predicts an absorbing vacuum phase ($\rho=0$) and an active phase density, with $\rho=0.37$, which contrasts with LIFE's absorbing states in a square lattice, which have a stationary density $\rho_{2D} \approx 0.03$.	
	Here, we study and classify mean-field maps for $6144$ outer-totalistic CA and compare them with the corresponding behavior found in the square lattice. 
	We show that the single-site mean-field approach gives qualitative (and even quantitative) predictions for most of them. 
	The transition region in rule space seems to correspond to a nonequilibrium discontinuous absorbing phase transition instead of a continuous order-disorder one. 
	We claim that LIFE is a quasicritical nucleation process where vacuum phase domains invade the alive phase.
	Therefore, LIFE is not at the ``border of chaos,'' but thrives on the ``border of extinction.''

\end{abstract}


\maketitle

\section{Introduction}

	The cellular automaton Game of LIFE (GL) \cite{ref14} had been extensively studied in the 1990s by statistical physicists.
	Bak, Chen and Creutz \cite{a1} claimed that LIFE is a system presenting self-organized criticality (SOC) without any conserved quantity \cite{ref18,r5}, while Bennett and Bourzutschky \cite{bennet} argued that the observed criticality was due to finite size effects.
	Since other nonconservative SOC models have also had their strict critical behavior contested \cite{PhysRevE.59.4964,1742-5468-2009-09-P09009}, several studies examined LIFE in detail, with the general conclusion that the GL is slightly subcritical \cite{a5,a6,a7,blok2}.
	Moreover, single-site mean-field approximations were developed for deducing GL densities, but no one could reproduce the numerical results from simulations in the square lattice \cite{woottersandlangton,McIntosh,cs2,cs1,Bagnoli1991}. 	
	Therefore, it was believed that mean-field approximations were not applicable to LIFE and were not very useful to  cellular automata (CA) rules in general.

	One decade before, Wolfram \cite{wolfram01,Wolfram02,ref16,wol7} proposed a qualitative classification for CA behavior. 
	This classification is composed of the \textit{Class I} (fixed point), \textit{Class II} (periodic), \textit{Class III} (chaotic) and \textit{Class IV} (``complex'') behaviors. 
	However, Wolfram's classes are only phenomenological descriptions: given a CA rule, it is not possible to predict to which class it pertains.
	An attempt in this predictive direction was made by Langton \cite{selflangton,Langton01}, who proposed the parameter $\lambda$ to classify the CA rules, which, unfortunately, failed at describing complex rules such as LIFE \cite{woottersandlangton, r2}.

	In this context, the questions that we want to explore are the following.
	In what sense is the GL critical (or subcritical)? 
	Is the single-site mean-field approximation not applicable to LIFE and other ``complex'' rules? 
	Is there any parameter for CA rule space (similar to a control parameter and obtained a \textit{priori} from the rule table) to order the CA rules and reveal any phase transition?
	What kind of phase transition is the GL related to?

	Our principal findings concern the usefulness of the single-site mean-field (MF) approximation. 
	We find that this kind of MF approximation can be applied to explain the density of live cells in the GL and to define a new control parameter.
	We describe LIFE behavior in terms of coexistence and competition between two phases and show that it corresponds to a subcritical (but quasicritical) nucleation process of living cells.

	For a large number of CA, we found that the MF predictions are qualitatively and even quantitatively correct.
	In the subspace of the $6144$ order $3$ rules (rules that have the MF equation dominated by $\rho^3$ when $\rho \rightarrow 0$), the MF analysis employed here predicts that $2203$ of them have only a trivial zero phase.
	For the remaining $3941$, the MF predicts a nontrivial phase $\rho^{*}$ which may not be stable to invasion by the zero phase when simulated in a square $(2D)$ lattice.
	These $3941$ rules includes $440$ automata that have period $T \geq 2$ and have been excluded from our study.
In the remaining $3501$ rules, most of the CA patterns found in a square lattice can be viewed as composed of vacuum and alive phase domains.

\section{The Model}

	We consider outer-totalistic binary bidimensional CA where each cell can assume the state $s=0$ (``dead'' or ``vacuum'') or $s=1$ (``alive'' or ``particle'').
	The update is made in parallel and realized according to the CA transition rule $R[h,s]$. 
	The rule $R[h,s]$ determines, for a given number $h(t)$ of alive neighbors and the state $s(t)$ of the central cell at time $t$, the next state $s(t+1)$ of the central cell. 
	The CA rule is called outer-totalistic because the rule does not depend on the exact neighbors configuration, but only on the total number of alive neighbors $h(t)$ and on the cell state $s(t)$.

	We use a Moore neighborhood with eight nearest neighbors.
	So, there are $2^{18}=262\:144$ different rules in the rule space.
	For LIFE, the transition rules are $R[0,3]=1$, $R[1,2]=1$ and $R[1,3]=1$, while all other configurations lead to a zero state at the next time step (see Table \ref{table1}).
	
	From now, we denote $\rho(t)$ as the mean-field density of alive cells, using $\rho_{MF}(t)$ when necessary to stress its origin. 
	Densities measured by simulations in the square lattice are denoted by $\rho_{2D}(t)$. 
	Our model was developed starting from the LIFE's rule table (Table \ref{table1}) and, by changing LIFE's rule, we also studied in detail $6143$ other order-3 cellular automata. 

\begin{table}[t!]
	\begin{center}
		\begin{tabular}{|c|c|c|}\hline
			$h$ & $s=0$ & $s=1$ \\ \hline
			$0$ & $0$ & $0$ \\ \hline
			$1$ & $0$ & $0$ \\ \hline
			$2$ & $0$ & $\textbf{1}$ \\ \hline
			$3$ & $\textbf{1}$ & $\textbf{1}$ \\ \hline
			$4$ & $\textbf{0}$ & $\textbf{0}$ \\ \hline
			$5$ & $\textbf{0}$ & $\textbf{0}$ \\ \hline
			$6$ & $\textbf{0}$ & $\textbf{0}$ \\ \hline
			$7$ & $\textbf{0}$ & $\textbf{0}$ \\ \hline
			$8$ & $\textbf{0}$ & $\textbf{0}$ \\ \hline
		\end{tabular}
	\end{center}
	\caption{LIFE's rule. According to the number of alive neighbors $h$ and the state $s$ of the central cell, the next state of the central cell will be given by the digits on the table. The bold ones were changed to generate the $6143$ other order-3 cellular automata rules also examined in this paper.}
	\label{table1}
\end{table}

\section{Mean-field calculations}

	In order to calculate an analytical density $\rho(t)$ of live cells for any CA rule, we use the single-site MF approximation. 
	In this approach, spatial correlations are neglected and one considers only the probability $P(s,h)$ of a site with state $s(t)$ to have $h(t)$ alive neighbors.
	The density $\rho(t+1)$ is written as

\begin{equation}
	\rho(t+1)=\sum^{1}_{s=0}\sum^{8}_{h=0}R[s,h]P_{t}(s,h).
	\label{eq1}
\end{equation}

	\noindent With a density $\rho(t)$, the probability of finding a live cell is $P_{t}(s=1)=\rho(t)$, while the probability of finding a dead cell is $P_{t}(s=0)=1-\rho(t)$. 
	Since no correlations are assumed between sites, the cell probability of having $h$ live neighbors follows a binomial distribution,

\begin{equation}
	P(h,t)=C^{8}_{h}\rho(t)^{h}(1-\rho(t))^{8-h},
	\label{eq2}
\end{equation}

	\noindent where $C^{8}_{h}$ is the binomial coefficient. The expression for $P_{t}(s,h)$ becomes:

\begin{equation}
	P_{t}(s,h)=s\rho(t)P(h,t)+(1-s)(1-\rho(t))P(h,t).
	\label{eq3}
\end{equation}

	This result implies that Eq.(\ref{eq1}) can be written as a map $\rho(t+1)=M(\rho(t))$:

\begin{eqnarray}
 M(\rho(t))= & (1-\rho(t))\sum^{8}_{h=0}R[0,h]P(h,t)+ \nonumber \\
  & \rho(t)\sum^{8}_{h=0}R[1,h]P(h,t). 	\label{eq4}
\end{eqnarray}

	This expression allows us to determine the mean-field return map for each CA rule and analyze its fixed points. 
	These maps are polynomials of order up to 9 in $\rho(t)$. 
	Applying LIFE's rule $R[s,h]$ to Eq. (\ref{eq4}), we obtain $\rho(t+1)=28\rho(t)^{3}(1-\rho(t))^{5}(3-\rho(t))$, which is shown in Fig.1 along with two other rules that present qualitatively different behaviors. 
	The rules are identified by Born/Survive nomenclature, $B$($h$ values that make a cell born)/$S$($h$ values to keep the cell alive).
	In this code, LIFE is the $B3/S23$ rule.

\begin{figure}[b]
	\centering
	\includegraphics[width=0.7\textwidth]{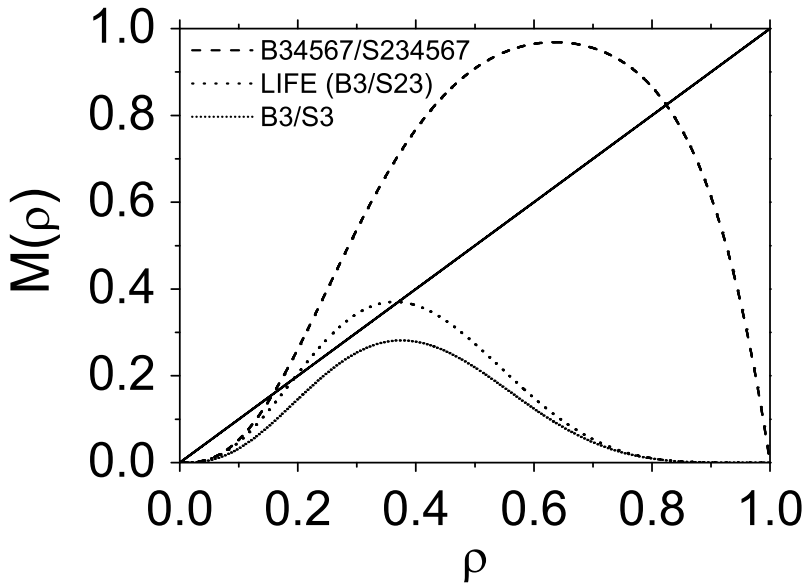}
	\label{fig1}
	\caption{MF return map generated by the application of specific rules to Eq. (\ref{eq4}). We see three different behaviors: rule $B34567/S234567$ has an absorbing state, a saddle point and a unstable fixed point (which leads to periodic behavior); LIFE has an absorbing state, a saddle point and a stable fixed point; and $B3/S3$ has only an absorbing state.}
\end{figure}

\section{Results}

	According to LIFE's return map, there are three fixed points. 
	Two of then, $\rho^{0}=0$ and $\rho^{*}=0.37$, are stable ones, while the fixed point $\rho^{s}=0.19$ is unstable (the saddle point). 
	These results are confirmed in simulations in a lattice where each cell has eight different random neighbors at each time step. 
	Indeed, the MF calculation reproduces well the behavior of any CA with random neighbors (quenched and annealed cases, not shown). 
	
	\begin{figure}[t]
	\centering
		\includegraphics[width=0.7\textwidth]{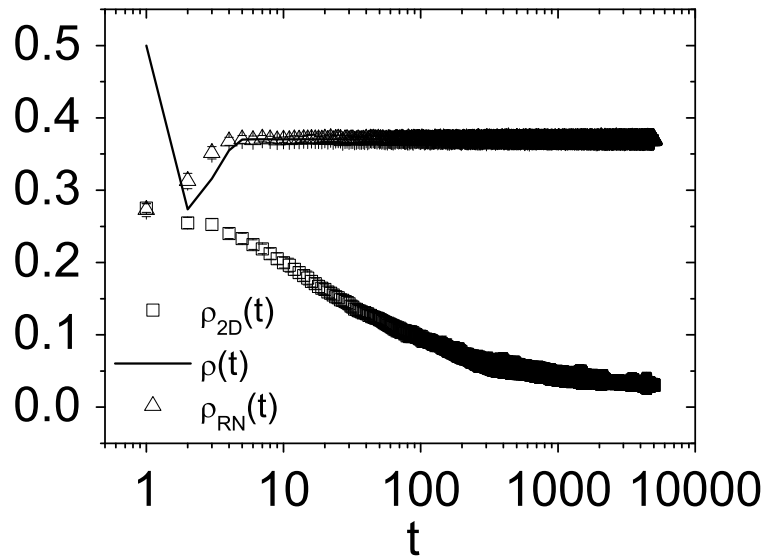}
	\caption{Density of LIFE live cells obtained by MF approximation $\rho$ (line), by simulations performed in a square lattice with Moore neighborhood $\rho_{2D}$, and by simulations in a lattice with eight random neighbors $\rho_{RN}$, with $N=L^{2}$ sites, $L=100$ and averages were performed over $30$ runs.}
	\label{fig2}
\end{figure}

	As mentioned before, the density from LIFE's simulations in a square lattice ($\rho_{2D}\approx0.03$) differs from the MF predictions ($\rho^*=0.37$), as seen in Fig. \ref{fig2}.
	This result is well known and here we give an explanation: suppose we put the $2D$ system in the initial condition $\rho_{2D}(t=0)=\rho^{*}$.
	Due to the initial density fluctuations, bubbles of vacuum phase appear and grow.
	This produces a lowering of $\rho_{2D}(t)$, which, at any time, is a spatial average of vacuum and $\rho^*$-like regions.
	
	Indeed, this occurs for generic random initial configurations.
	For special initial conditions, we can construct metastable states of higher densities.
	As an example, the most compact state created with blocks (a stable LIFE's structure that is a square composed by $2\times2$ cells in the lattice) separated by lines presents a metastable density of $\rho_{2D}=\frac{4}{9}=0.444\dots$ .

\begin{figure}[b]
	\centering
	\includegraphics[width=0.7\textwidth]{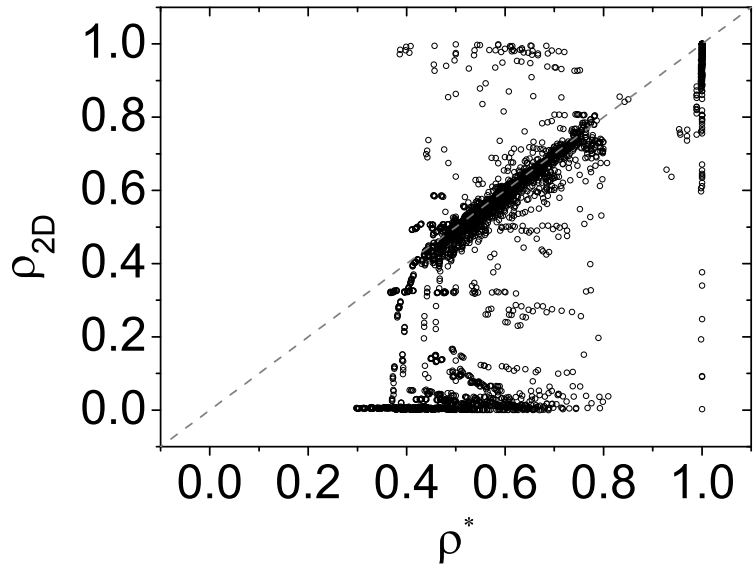}
	\caption{Relation between densities from lattice measurements ($\rho_{2D}$) and the single-site approximation ($\rho=\rho^{*}$). The initial condition in the simulations is $\rho_{2D}(t=0)=\rho^{*}$. The lattice size is $L=100$ and the averages were performed over $30$ runs.}
	\label{fig3}
\end{figure}

	In $2D$ simulations we used the MF stable fixed point $\rho^*$ as the initial condition for the correspondent rule.
	In Fig. \ref{fig3} we plot $\rho_{2D}$ versus its corresponding $\rho^*$ for the $3501$ rules where $\rho^*$ is stable. 
	We see that a large number of rules can have its stationary densities ($\rho_{2D}$) estimated by the single-site approximation (points around the line $\rho_{2D}=\rho^*$).
	In these rules cases, the entire lattice is dominated by a single homogeneous phase whose density is correlated with the non-zero MF return map stable fixed point $\rho^{*}$.

	Actually, this homogeneous $\rho_{2D}$ phase is not exactly the same as the mean-field phase $\rho^{*}$, since there is spatial correlations in the $2D$ lattice.
	However, the fact that $\rho_{2D} \approx \rho^{*}$ and that the probability $P(h,t)$ of a cell to have $h$ live neighbors is $P_{2D}(h,t) \approx P_{MF}(h,t)$ (see Fig. \ref{fig40}) seem to indicate that these correlations are weak.
	
	In Fig. \ref{fig3} we also observe CA where the initial condition is not stable and decays to the zero phase, $\rho_{2D}(t\rightarrow\infty)\approx\rho^{0}$. 
	Since $\rho^{*}$ is stable for random neighbor lattices, we presume that the square lattice allows, due to fluctuations, the formation of bubbles (or nuclei) of zero phase and that this nucleation process enables the zero phase to expand and overcome the $\rho^{*}$ phase.

	We also find CA where a coexistence of vacuum (with $\rho^{0}=0$) and alive (with $\rho\approx \rho^{*}$) domains is achieved, meaning that $\rho^{0}<\rho_{2D}<\rho^{*}$. 
	In these cases, we can describe $\rho_{2D}(t)$ as a linear combination 

\begin{equation}
	\rho_{2D}(t)\approx A^{*}(t)\rho^{*}+A^{0}(t)\rho^{0}+A^{+}(t)\rho^{+}.
	\label{eq5}
\end{equation}

	The terms $A^{*}(t)$, $A^{0}(t)$ and $A^{+}(t)$ are related to  the fraction of regions (or areas) with densities $\rho^{*}$, $\rho^{0}$ and $\rho^{+}$, respectively.
	
\begin{figure}[t]
	\centering
	\includegraphics[width=0.5\textwidth]{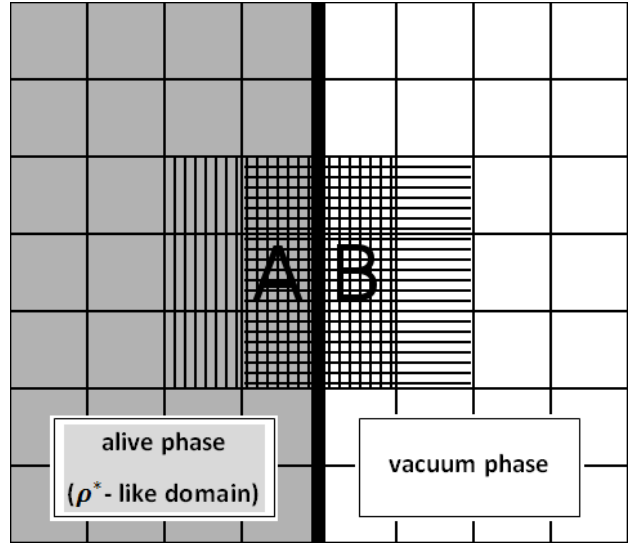}
	\caption{Site A belongs to the gray region, which represents the region with density  $\rho^*$, while site B is in the white region, which represents the region with density $\rho^0$. The thick black line is the interface between these two regions. The vertical and horizontal lines determine the neighborhood of the sites A and B, respectively.}
	\label{borderfig}
\end{figure}

	Densities $\rho^{*}$ and $\rho^{0}$ are stable fixed points that come from the approximation that the $2D$ alive phase has density $\rho^*$.
	We call $\rho^+$ the interfarcial density, which plays a crucial role.
	From Fig. \ref{borderfig}, we obtain that, for large bubbles (corresponding to linear interfaces), we can approximate the interfacial density in the neighborhood of $A$ ($\rho_A$) and $B$ ($\rho_B$) as
	
	\[ \rho^+ \approx \frac{1}{2} \left[ \rho_A + \rho_B \right]= \frac{1}{2} \left[ \left( \frac{6}{9}\rho^* + \frac{3}{9}\rho^0 \right) + \left(  \frac{3}{9}\rho^* + \frac{6}{9}\rho^0 \right) \right] = 	
	\frac{\rho^* + \rho^0}{2} = \frac{\rho^*}{2} . \]

	Notice that the density $\rho_{2D}$ given by Eq. (\ref{eq5}) is also valid for the transient regime, and not only for the stationary density. 
	The time dependence appears in the evolution of the coefficients $A^{*}(t)$, $A^{0}(t)$ and $A^{+}(t)$. 
	We note that there is only two free coefficients, since $A^{*}(t)+A^{0}(t)+A^{+}(t)=1$.

	It is important to stress that we have observed that the neighbor probability $P_{2D}(h,t)$ also can be fitted as a sum (Fig. \ref{fig40}):
	
	\begin{figure}[b]
	\centering
		\subfigure[]{\includegraphics[width=0.45\textwidth]{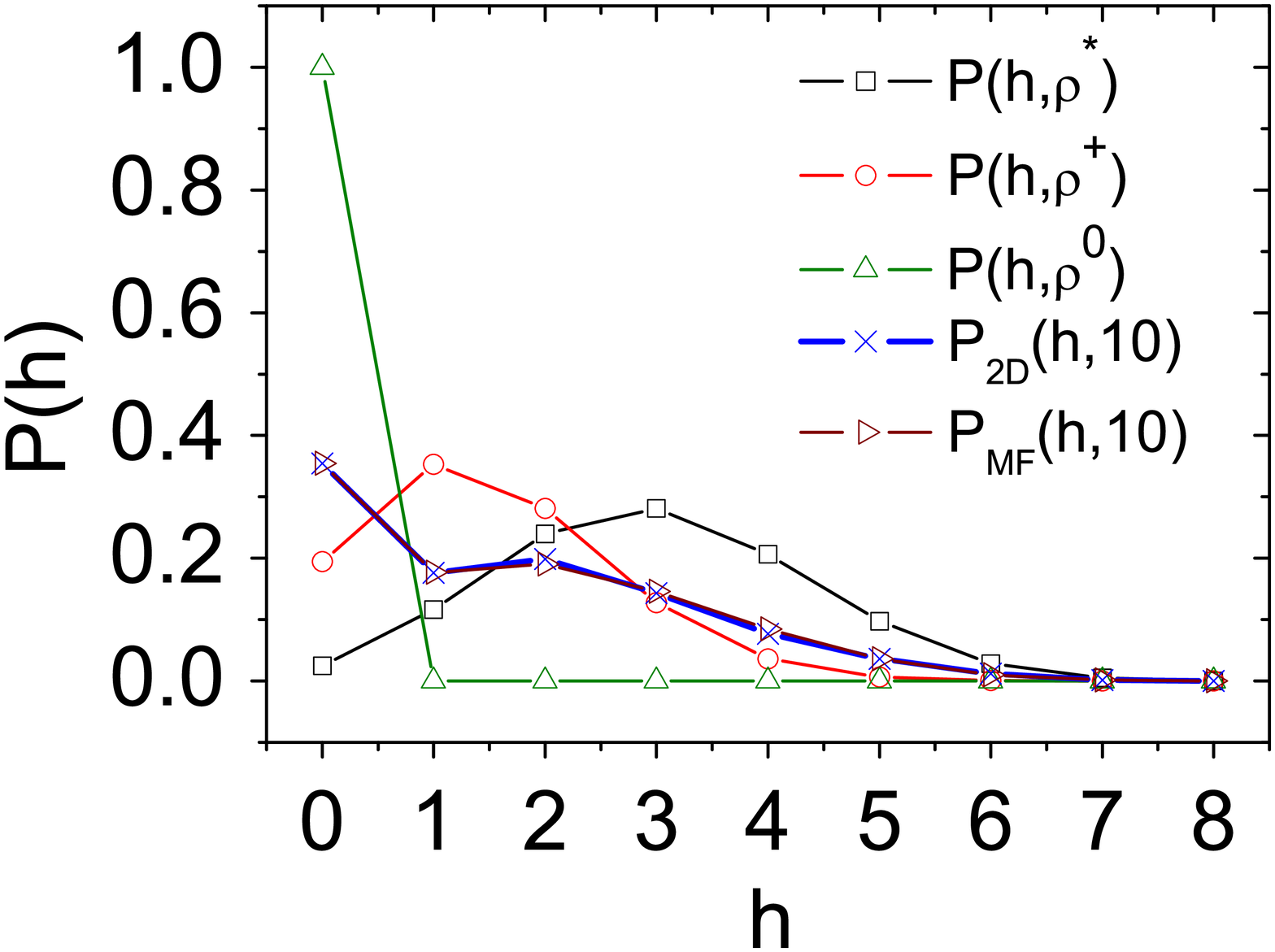}}
		\subfigure[]{\includegraphics[width=0.45\textwidth]{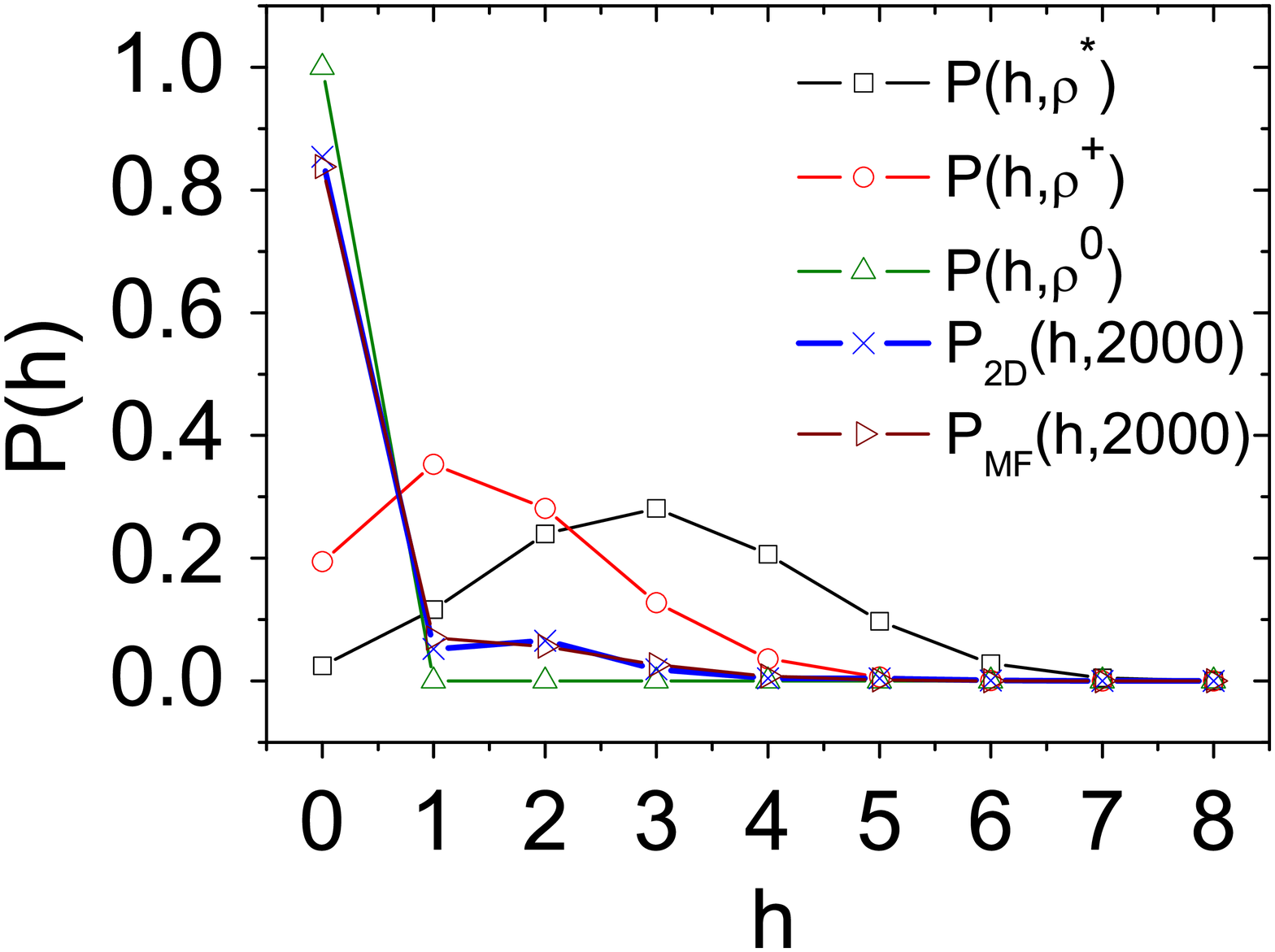}}
		\caption{Probabilities $P(h,\rho^{*})$, $P(h,\rho^{+})$, $P(h,\rho^{0})$, $P_{2D}(h,t)$ and $P_{MF}(h,t)$, from Eq. (\ref{eq6}), for different times in LIFE. In (a) the values of coefficients are: $A^{*}(10)=0.34$, $A^{+}(10)=0.39$ and $A^{0}(10)=0.27$. In (b), we have $A^{*}(2000)=0.01$, $A^{+}(2000)=0.17$ and $A^{0}(2000)=0.82$. The $P_{2D}$ and the $P_{MF}$ curves are almost indistinguishable. The fit only works if we use the interfacial term $P(h,\rho^+)$.}
	\label{fig40}
\end{figure}

\begin{eqnarray}
	P_{2D}(h,t) \approx & A^{*}(t)P(h,\rho^{*})+A^{0}(t)P(h,\rho^{0}) \nonumber \\
	& +A^{+}(t)P(h,\rho^{+}) =  P_{MF}(h,t). 	\label{eq6}
\end{eqnarray}

	Following this heuristic scenario where bubbles of zero phase invade the $\rho^{*}$ phase, we propose a ``control parameter'' for these CA. 
	We notice that, if the bulk densities $\rho^{0}$ and $\rho^{*}$ are stable, then the zero phase can grow mostly at the interfaces. 
	The density of zero sites is $1-\rho(t)$ at a given time $t$ and it grows to $1-\rho(t+1)=1-M(\rho)$ at the next time step. 
	So, we define the growth rate for zero sites at interfaces as

\begin{equation}
	\sigma_{0}=\frac{1-M(\rho^{+})}{1-\rho^{+}}.
	\label{eq7}
\end{equation}

	This means that, if $\sigma_{0}>1$, the zero phase expands and, if $\sigma_{0}<1$, the zero phase contracts. 
	The critical growth is $\sigma_{0}=1$.
	Notice that the parameter $\sigma_{0}$ is heuristic and MF-like. 
	Remembering that $\rho^{+}=\rho^{*}/2$, we calculate $\sigma_{0}=[1-M(\rho^{*}/2)][1-\rho^{*}/2]$ by using the MF value for $\rho^{*}$ and the return map $M(\rho)$, which means that $\sigma_{0}$ is a parameter calculable \textit{a priori} from the rule table. 
	
	A plot with the control parameter $\sigma_{0}$ is given in Fig. \ref{fig4}. 
	If the rule has its density estimated by the MF approximation, the order parameter $\rho'= \rho_{2D} / \rho^*$ is close to one and the point lies around the line $\rho'=1$. 
	If the order parameter $\rho'\neq 1$, then the behavior can be distinguished in the following three cases:
	
	\begin{figure}[!b!]
	\centering
	\includegraphics[width=0.7\textwidth]{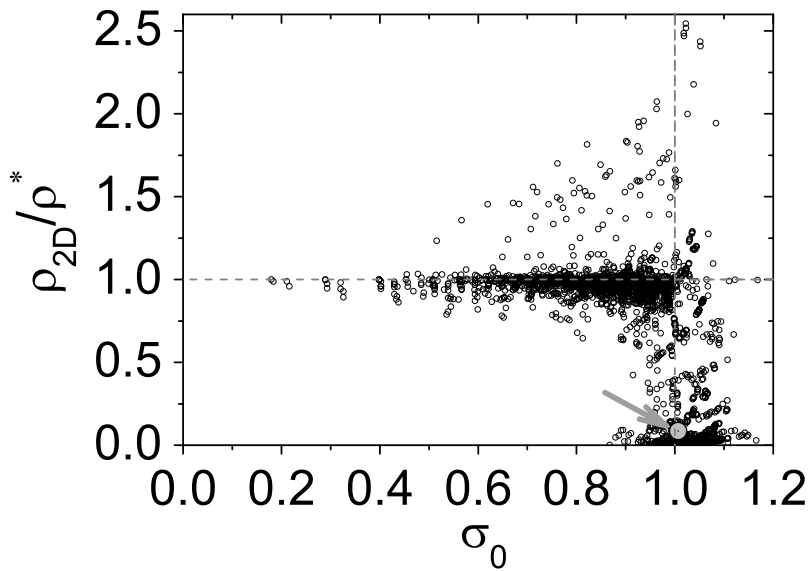}
	\caption{Relation between $\rho'=\rho_{2D} / \rho^*$ and $\sigma_{0}$. Results were obtained for $L=100$ and averaged over $30$ runs. The gray arrow points the LIFE's position (large gray dot).}
	\label{fig4}
\end{figure}

\begin{enumerate}[I)]
	\item $\rho'>1$. Correlations in the $2D$ lattice promote an over-activity. The CA present a high density $\rho_{2D}$, with strong spatial correlation that can not be approximated by Eqs. (\ref{eq5}) and (\ref{eq6});
	\item $1<\rho'<0$. The MF approximation overestimates the density of live cells. By examining several CA we find that $1<\rho'<0$ indicates a kind of coexistence between domains of the $\rho^{*}$-like and the zero phase;
	\item $\rho'=0$. The activity of the lattice is driven to a stable absorbing state not expected by the MF approximation [for the initial density $\rho_{2D}(t=0)=\rho^*$]. That is, the zero phase invades and eliminates the $\rho^{*}$-like phase in $D=2$.
\end{enumerate}

	The most interesting behaviors are provided by the case II rules (which includes LIFE). 
	These rules can be described by Eq. (\ref{eq5}) and can be interpreted as a mixture of the vacuum $\rho=0$ and $\rho \approx \rho^{*}$ phases. 
	Notice that LIFE ($\sigma_{0} = 1.006$, see Fig. \ref{fig4}) is near criticality in the sense that zero phase nucleation is slow (a power law growth), almost eliminating the $\rho^*$ domains.
	Figure \ref{fig4} suggests that the most relevant phase transition in our CA rule space is a first-order absorbing transition, not a second-order transition as conjectured by some authors \cite{r2,Langton1990}.

\section{Conclusion}

	In this paper, we show that several single-site MF results are useful to provide qualitative and even quantitative understandings of CAs in $2D$ lattices.
	In particular, LIFE is a special case where the vacuum phase is slightly super critical ($\sigma_{0}=1.006$) or, for the alive phase, the nucleation process is slightly subcritical. 
	Furthermore, the complex behavior of GL seems indeed to be related to a phase transition in CA rule space which reminds us a first-order absorbing phase transition with metastable states.

	With respect to LIFE, Bagnoli \textit{et al.} \cite{Bagnoli1991} implemented a high-order MF calculation to capture temporal correlations.
	Specifically, the MF map is extended to time $t=2$ and compared to MF from $t=1$, that corresponds to  Eq. (\ref{eq1}) (for further details, see \cite{Bagnoli1991}).	
	However, their approach seems to be insufficient to take into account the fact that the $\rho_{2D}$ in the square lattice refers to a spatial average of domains with zero density and domains with high density (near 0.37). 
	They also proposed an interesting model of deposition of animals (disks) with removal in case of collisions, which predicts well the stationary density, but it is not clear if such a uniform deposition model can reproduce the presence of large regions with zero density obtained by the direct simulation of LIFE.

	In conclusion, MF results for $\rho^{*}$ give a good approximation to $\rho_{2D}$ (and also detect special CA where $\rho_{2D}>\rho^{*}$). 
	The $ \rho_{2D}<\rho^{*} $ cases seem to correspond to metastable mixture states between the vacuum and a $\rho^{*}$-like phase.
	In this sense, LIFE is a fine-tuned quasicritical nucleation process at the border of extinction.	
	
	Curiously, a similar result was found recently by Degrassi \textit{et al.} \cite{Degrassi} and by Buttazzo \textit{et al.} \cite{Buttazzo} concerning the vacuum stability in the Standard Model.
	They have found that our universe seems to be in the quasicritical metastable region and conjecture that this occurs due to self-organized criticality.
	However, if an analogy between the CA rule space and the multiverse rule space were made, we would see that complex automata are rare, tending to a null measure as this space grows. 
	LIFE's rule, with its rare property of being an Universal Turing Machine (UTM) \cite{Rendell}, is fine-tuned to place the CA at the border of the phase transition ($\sigma_0 = 1.006$).
	Similar to LIFE, our Universe is also a UTM, and perhaps its near-critical vacuum state is related to class IV complex behavior. 
	However, if we desire that complex automata would be attractors in the CA rule space, some dynamics in the rule table must be proposed (for example, mutation and selection of CAs with larger relaxation times).

\section{Acknowledgments}
	S. Reia CAPES for the financial support and Ariadne A. Costa for useful conversations.
	O. Kinouchi acknowledges support from CNPq and CNAIPS-USP.

\bibliographystyle{apsrev}
\bibliography{references}

\begin{thebibliography}{27}
\expandafter\ifx\csname natexlab\endcsname\relax\def\natexlab#1{#1}\fi
\expandafter\ifx\csname bibnamefont\endcsname\relax
  \def\bibnamefont#1{#1}\fi
\expandafter\ifx\csname bibfnamefont\endcsname\relax
  \def\bibfnamefont#1{#1}\fi
\expandafter\ifx\csname citenamefont\endcsname\relax
  \def\citenamefont#1{#1}\fi
\expandafter\ifx\csname url\endcsname\relax
  \def\url#1{\texttt{#1}}\fi
\expandafter\ifx\csname urlprefix\endcsname\relax\def\urlprefix{URL }\fi
\providecommand{\bibinfo}[2]{#2}
\providecommand{\eprint}[2][]{\url{#2}}

\bibitem[{\citenamefont{Berlekamp et~al.}(1982)\citenamefont{Berlekamp, Conway,
  and Guy}}]{ref14}
\bibinfo{author}{\bibfnamefont{E.~R.} \bibnamefont{Berlekamp}},
  \bibinfo{author}{\bibfnamefont{J.~H.} \bibnamefont{Conway}},
  \bibnamefont{and} \bibinfo{author}{\bibfnamefont{R.~K.} \bibnamefont{Guy}},
  \emph{\bibinfo{title}{Winning Ways for Your Mathematical Plays}},
  vol.~\bibinfo{volume}{2} (\bibinfo{publisher}{Peters, Natick, Massachussets},
  \bibinfo{year}{1982}).

\bibitem[{\citenamefont{Bak et~al.}(1989)\citenamefont{Bak, Chen, and
  Creutz}}]{a1}
\bibinfo{author}{\bibfnamefont{P.}~\bibnamefont{Bak}},
  \bibinfo{author}{\bibfnamefont{K.}~\bibnamefont{Chen}}, \bibnamefont{and}
  \bibinfo{author}{\bibfnamefont{M.}~\bibnamefont{Creutz}},
  \bibinfo{journal}{Nature (London)} \textbf{\bibinfo{volume}{342}},
  \bibinfo{pages}{780} (\bibinfo{year}{1989}).

\bibitem[{\citenamefont{Bak et~al.}(1987)\citenamefont{Bak, Tang, and
  Wiesenfeld}}]{ref18}
\bibinfo{author}{\bibfnamefont{P.}~\bibnamefont{Bak}},
  \bibinfo{author}{\bibfnamefont{C.}~\bibnamefont{Tang}}, \bibnamefont{and}
  \bibinfo{author}{\bibfnamefont{K.}~\bibnamefont{Wiesenfeld}},
  \bibinfo{journal}{Phys. Rev. Lett.} \textbf{\bibinfo{volume}{59}},
  \bibinfo{pages}{381} (\bibinfo{year}{1987}).

\bibitem[{\citenamefont{Bak}(1992)}]{r5}
\bibinfo{author}{\bibfnamefont{P.}~\bibnamefont{Bak}}, \bibinfo{journal}{Phys.
  A (Amsterdam, Neth.)} \textbf{\bibinfo{volume}{191}}, \bibinfo{pages}{41}
  (\bibinfo{year}{1992}).

\bibitem[{\citenamefont{Bennett and Bourzutschky}(1991)}]{bennet}
\bibinfo{author}{\bibfnamefont{C.}~\bibnamefont{Bennett}} \bibnamefont{and}
  \bibinfo{author}{\bibfnamefont{M.~S.} \bibnamefont{Bourzutschky}},
  \bibinfo{journal}{Nature (London)} \textbf{\bibinfo{volume}{350}},
  \bibinfo{pages}{468} (\bibinfo{year}{1991}).

\bibitem[{\citenamefont{Kinouchi and Prado}(1999)}]{PhysRevE.59.4964}
\bibinfo{author}{\bibfnamefont{O.}~\bibnamefont{Kinouchi}} \bibnamefont{and}
  \bibinfo{author}{\bibfnamefont{C.~P.~C.} \bibnamefont{Prado}},
  \bibinfo{journal}{Phys. Rev. E} \textbf{\bibinfo{volume}{59}},
  \bibinfo{pages}{4964} (\bibinfo{year}{1999}).

\bibitem[{\citenamefont{Bonachela and Muñoz}(2009)}]{1742-5468-2009-09-P09009}
\bibinfo{author}{\bibfnamefont{J.~A.} \bibnamefont{Bonachela}}
  \bibnamefont{and} \bibinfo{author}{\bibfnamefont{M.~A.} \bibnamefont{Muñoz}},
  \bibinfo{journal}{J. Stat. Mech.} \bibinfo{eid}{P09009}
  (\bibinfo{year}{2009}).

\bibitem[{\citenamefont{Garcia et~al.}(1993)\citenamefont{Garcia, Gomes, Jyh,
  Ren, and Sales}}]{a5}
\bibinfo{author}{\bibfnamefont{J.~B.~C.} \bibnamefont{Garcia}},
  \bibinfo{author}{\bibfnamefont{M.~A.~F.} \bibnamefont{Gomes}},
  \bibinfo{author}{\bibfnamefont{T.~I.} \bibnamefont{Jyh}},
  \bibinfo{author}{\bibfnamefont{T.~I.} \bibnamefont{Ren}}, \bibnamefont{and}
  \bibinfo{author}{\bibfnamefont{T.~R.~M.} \bibnamefont{Sales}},
  \bibinfo{journal}{Phys. Rev. E} \textbf{\bibinfo{volume}{48}},
  \bibinfo{pages}{3345} (\bibinfo{year}{1993}).

\bibitem[{\citenamefont{Alstr\o{}m and Leao}(1994)}]{a6}
\bibinfo{author}{\bibfnamefont{P.}~\bibnamefont{Alstr\o{}m}} \bibnamefont{and}
  \bibinfo{author}{\bibfnamefont{J.}~\bibnamefont{Leao}},
  \bibinfo{journal}{Phys. Rev. E} \textbf{\bibinfo{volume}{49}},
  \bibinfo{pages}{R2507} (\bibinfo{year}{1994}).

\bibitem[{\citenamefont{Hemmingsson}(1995)}]{a7}
\bibinfo{author}{\bibfnamefont{J.}~\bibnamefont{Hemmingsson}},
  \bibinfo{journal}{Phys. D (Amsterdam, Neth.)} \textbf{\bibinfo{volume}{80}},
  \bibinfo{pages}{151 } (\bibinfo{year}{1995}).

\bibitem[{\citenamefont{Blok and Bergersen}(1997)}]{blok2}
\bibinfo{author}{\bibfnamefont{H.~J.} \bibnamefont{Blok}} \bibnamefont{and}
  \bibinfo{author}{\bibfnamefont{B.}~\bibnamefont{Bergersen}},
  \bibinfo{journal}{Phys. Rev. E} \textbf{\bibinfo{volume}{55}},
  \bibinfo{pages}{6249} (\bibinfo{year}{1997}).

\bibitem[{\citenamefont{Wootters and Langton}(1990)}]{woottersandlangton}
\bibinfo{author}{\bibfnamefont{W.~K.} \bibnamefont{Wootters}} \bibnamefont{and}
  \bibinfo{author}{\bibfnamefont{C.~G.} \bibnamefont{Langton}},
  \bibinfo{journal}{Phys. D (Amsterdam, Neth.)} \textbf{\bibinfo{volume}{45}},
  \bibinfo{pages}{95 } (\bibinfo{year}{1990}).

\bibitem[{\citenamefont{McIntosh}(1990)}]{McIntosh}
\bibinfo{author}{\bibfnamefont{H.~V.} \bibnamefont{McIntosh}},
  \bibinfo{journal}{Phys. D (Amsterdam, Neth.)} \textbf{\bibinfo{volume}{45}},
  \bibinfo{pages}{105 } (\bibinfo{year}{1990}).

\bibitem[{\citenamefont{Gutowitz and Victor}(1989)}]{cs2}
\bibinfo{author}{\bibfnamefont{H.~A.} \bibnamefont{Gutowitz}} \bibnamefont{and}
  \bibinfo{author}{\bibfnamefont{J.~D.} \bibnamefont{Victor}},
  \bibinfo{journal}{J. Stat. Phys.} \textbf{\bibinfo{volume}{54}},
  \bibinfo{pages}{495} (\bibinfo{year}{1989}).

\bibitem[{\citenamefont{Gutowitz and Victor}(1987)}]{cs1}
\bibinfo{author}{\bibfnamefont{H.~A.} \bibnamefont{Gutowitz}} \bibnamefont{and}
  \bibinfo{author}{\bibfnamefont{J.~D.} \bibnamefont{Victor}},
  \bibinfo{journal}{Complex Systems} \textbf{\bibinfo{volume}{1}},
  \bibinfo{pages}{57} (\bibinfo{year}{1987}).

\bibitem[{\citenamefont{Bagnoli et~al.}(1991)\citenamefont{Bagnoli, Rechtman,
  and Ruffo}}]{Bagnoli1991}
\bibinfo{author}{\bibfnamefont{F.}~\bibnamefont{Bagnoli}},
  \bibinfo{author}{\bibfnamefont{R.}~\bibnamefont{Rechtman}}, \bibnamefont{and}
  \bibinfo{author}{\bibfnamefont{S.}~\bibnamefont{Ruffo}},
  \bibinfo{journal}{Phys. A (Amsterdam, Neth.)} \textbf{\bibinfo{volume}{171}},
  \bibinfo{pages}{249 } (\bibinfo{year}{1991}).

\bibitem[{\citenamefont{Wolfram}(1983)}]{wolfram01}
\bibinfo{author}{\bibfnamefont{S.}~\bibnamefont{Wolfram}},
  \bibinfo{journal}{Rev. Mod. Phys.} \textbf{\bibinfo{volume}{55}},
  \bibinfo{pages}{601} (\bibinfo{year}{1983}).

\bibitem[{\citenamefont{Wolfram}(1984{\natexlab{a}})}]{Wolfram02}
\bibinfo{author}{\bibfnamefont{S.}~\bibnamefont{Wolfram}},
  \bibinfo{journal}{Phys. D (Amsterdam, Neth.)} \textbf{\bibinfo{volume}{10}},
  \bibinfo{pages}{1} (\bibinfo{year}{1984}{\natexlab{a}}).

\bibitem[{\citenamefont{Wolfram}(1984{\natexlab{b}})}]{ref16}
\bibinfo{author}{\bibfnamefont{S.}~\bibnamefont{Wolfram}},
  \bibinfo{journal}{Nature (London)} \textbf{\bibinfo{volume}{311}},
  \bibinfo{pages}{419} (\bibinfo{year}{1984}{\natexlab{b}}).

\bibitem[{\citenamefont{Wolfram}(1986)}]{wol7}
\bibinfo{author}{\bibfnamefont{S.}~\bibnamefont{Wolfram}},
  \emph{\bibinfo{title}{Theory and applications of cellular automata}},
  Advanced Series on Complex Systems (\bibinfo{publisher}{World Scientific,
  Singapore}, \bibinfo{year}{1986}).

\bibitem[{\citenamefont{Langton}(1984)}]{selflangton}
\bibinfo{author}{\bibfnamefont{C.~G.} \bibnamefont{Langton}},
  \bibinfo{journal}{Phys. D (Amsterdam, Neth.)} \textbf{\bibinfo{volume}{10}},
  \bibinfo{pages}{135} (\bibinfo{year}{1984}).

\bibitem[{\citenamefont{Langton}(1986)}]{Langton01}
\bibinfo{author}{\bibfnamefont{C.~G.} \bibnamefont{Langton}},
  \bibinfo{journal}{Phys. D (Amsterdam, Neth.)} \textbf{\bibinfo{volume}{22}},
  \bibinfo{pages}{120 } (\bibinfo{year}{1986}).

\bibitem[{\citenamefont{Li et~al.}(1990)\citenamefont{Li, Packard, and
  Langton}}]{r2}
\bibinfo{author}{\bibfnamefont{W.}~\bibnamefont{Li}},
  \bibinfo{author}{\bibfnamefont{N.~H.} \bibnamefont{Packard}},
  \bibnamefont{and} \bibinfo{author}{\bibfnamefont{C.~G.}
  \bibnamefont{Langton}}, \bibinfo{journal}{Phys. D (Amsterdam, Neth.)}
  \textbf{\bibinfo{volume}{45}}, \bibinfo{pages}{77 } (\bibinfo{year}{1990}).

\bibitem[{\citenamefont{Langton}(1990)}]{Langton1990}
\bibinfo{author}{\bibfnamefont{C.~G.} \bibnamefont{Langton}},
  \bibinfo{journal}{Phys. D (Amsterdam, Neth.)} \textbf{\bibinfo{volume}{42}},
  \bibinfo{pages}{12} (\bibinfo{year}{1990}).

\bibitem[{\citenamefont{Degrassi et~al.}(2012)\citenamefont{Degrassi, Di~Vita,
  Elias-Miró, Espinosa, Giudice, Isidori, and Strumia}}]{Degrassi}
\bibinfo{author}{\bibfnamefont{G.}~\bibnamefont{Degrassi}},
  \bibinfo{author}{\bibfnamefont{S.}~\bibnamefont{Di~Vita}},
  \bibinfo{author}{\bibfnamefont{J.}~\bibnamefont{Elias-Miró}},
  \bibinfo{author}{\bibfnamefont{J.~R.} \bibnamefont{Espinosa}},
  \bibinfo{author}{\bibfnamefont{G.~F.} \bibnamefont{Giudice}},
  \bibinfo{author}{\bibfnamefont{G.}~\bibnamefont{Isidori}}, \bibnamefont{and}
  \bibinfo{author}{\bibfnamefont{A.}~\bibnamefont{Strumia}},
  \bibinfo{journal}{Journal of High Energy Phys.} \bibinfo{eid}{98}
  (\bibinfo{year}{2012}).

\bibitem[{\citenamefont{Buttazzo et~al.}(2013)\citenamefont{Buttazzo, Degrassi,
  Giardino, Giudice, Sala, Salvio, and Strumia}}]{Buttazzo}
\bibinfo{author}{\bibfnamefont{D.}~\bibnamefont{Buttazzo}},
  \bibinfo{author}{\bibfnamefont{G.}~\bibnamefont{Degrassi}},
  \bibinfo{author}{\bibfnamefont{P.}~\bibnamefont{Giardino}},
  \bibinfo{author}{\bibfnamefont{G.}~\bibnamefont{Giudice}},
  \bibinfo{author}{\bibfnamefont{F.}~\bibnamefont{Sala}},
  \bibinfo{author}{\bibfnamefont{A.}~\bibnamefont{Salvio}}, \bibnamefont{and}
  \bibinfo{author}{\bibfnamefont{A.}~\bibnamefont{Strumia}},
  \bibinfo{journal}{Journal of High Energy Phys.} \bibinfo{eid}{89}
  (\bibinfo{year}{2013}).

\bibitem[{\citenamefont{Rendell}(2011)}]{Rendell}
\bibinfo{author}{\bibfnamefont{P.}~\bibnamefont{Rendell}}, in
  \emph{\bibinfo{booktitle}{2011 International Conference on High Performance
  Computing and Simulation (HPCS)}} (\bibinfo{organization}{IEEE, Piscataway,
  NJ}, \bibinfo{year}{2011}), pp. \bibinfo{pages}{764--772}.

\end{thebibliography}

\end{document}